\documentclass[ aps,%
12pt,%
floatfix,%
final,%
notitlepage,%
oneside,%
onecolumn,%
nobibnotes,%
nofootinbib,%
superscriptaddress,%
showpacs,%
]%
{revtex4}

\begin{document}

\title{$c$-quark decay modes in $B_c$-meson}
\author{S.S. Gershtein} \email{Semen.Gershtein@ihep.ru}
\author{A.K. Likhoded} \email{Likhoded@ihep.ru}
\affiliation{Institute for High Energy Physics, Protvino, Russia}

\begin{abstract}
We discuss the possibility for observing $B_c$ mesons in decay channels with $B_s$ in final state.
\end{abstract}

\maketitle

\section{Introduction}

$B_c$-meson is the heaviest of the stable under strong interaction mesons. Because of its unique properties the study of the processes of its production and decay can be used to check current models of quark dynamics. Since both constituent quarks in this meson are heavy, one can use perturbation QCD for the calculation of $B_c$ production cross section. Very interesting also is the study of $B_c$ decay channels. Since decay of $B_c$ meson proceeds mostly (in 70\% of cases) through the decay of $c$-quark, one can expect a large yield of $B_s$ mesons, that is a significant part of directly produced $B_s$, since the cross section of the last process is an order of magnitude smaller then the cross section of $B_u$ and $B_d$ production. In our paper we would like to note, that the large number of $B_s$-mesons, used to study $B_s$-$\bar B_s$ oscillations at CDF \cite{R1} and D0 \cite{R2}, can also help to detect a noticeable amount of $B_c$-mesons in their decay to $B_s$. We also draw an attention to the surprisingly large experimental ratio of $B_c$ to $B_u$ production cross sections  \cite{R6}, that is an order of magnitude higher, than the theoretical predictions. So a thorough investigation of this problem is required.

\section{Mass spectrum}

The ground state of $(\bar b c)$ system is intermediate between charmonium and bottomonium. Since $B_c$ meson have both charm and bottom flavors opened, it gives an opportunity to study the dynamics of heavy quarks in addition to $(c\bar c)$ and $(b\bar b)$ systems. There are 16 narrow $(\bar b c)$ states below the threshold of $\bar B D$-pair production. In contrary to $(c\bar c)$ and $(b\bar b)$-systems, there are no annihilation channel for $(\bar b c)$-meson decay, so excited states can decay only to the ground states with the emission of photons and $\pi$-mesons.

The most accurate prediction of the masses of $B_c$-mesons (including excited states) were obtained in the framework of nonrelativistic potential models, that is based on NRQCD expansion in inverse quark mass $m_Q^{-1}$ and the relative quark velocity $v\to 0$ \cite{R3}. The errors of these predictions are $\sim 30$ MeV. In addition to potential models, the mass of the ground $(\bar b c)$-state was estimated also with the help of QCD sum rules and lattice QCD \cite{R4}. The results of these estimates are in good agreement with the experimental value $m_{B_c}=6276.5 \pm 4.0 \pm 2.7\,\mathrm{MeV/c^2}$, that was measured recently by CDF collaboration in fully reconstructed exclusive decay $B_c\to J/\psi\pi$ \cite{R5}.

\section{Dominant decay modes}

Both $B_c\to J/\psi \pi$ and semileptonic $B_c$ decay correspond to the transition $b\to c$. Semileptonic decay mode was used recently by CDF and D0 collaborations to measure the lifetime of $B_c$-meson \cite{R6}:
\begin{eqnarray*}
\tau_{B_c} &=& 0.448^{+0.123}_{-0.096}\,\pm 0.121\mathrm{ps}.
\end{eqnarray*}
This value is in a good agreement with theoretical calculations, that were based on the inclusive approach and the sum of dominant exclusive decay modes \cite{R7}:
\begin{eqnarray*}
\tau_{B_c}&=&0.48\pm 0.015\,\mbox{ps}.
\end{eqnarray*}
In table \ref{tab} we give the predictions of the branching fractions of exclusive $B_c$ decays, that were obtained in the framework operator product expansion (OPE), potential models and QCD sum rules. The main contribution to $B_c$ lifetime are caused by the decay of $c$-quark (70\%), while the contributions of $b$-quark decay and weak annihilation are $20\%$ and $10\%$ respectively.

While estimating the exclusive decay width in the framework of QCD sum rules, for the calculation of form factors it is important to take $\alpha_s/v$ corrections into account. These from factors satisfy well the relations, obtained with the help of NRQCD spin symmetry and effective theory of heavy quarks (HQET).

\begin{table}
\begin{center}
\begin{tabular}{|l|r|}
\hline
~~~~~Mode & BR, \%\\
\hline
 $B_c^+ \rightarrow \eta_c e^+ \nu$
 & 0.75\\
 $B_c^+ \rightarrow \eta_c \tau^+ \nu$
 & 0.23\\
 $B_c^+ \rightarrow \eta_c^\prime e^+ \nu$
 & 0.041\\
 $B_c^+ \rightarrow \eta_c^\prime \tau^+ \nu$
 & 0.0034\\
 $B_c^+ \rightarrow J/\psi e^+ \nu $
 & 1.9\\
 $B_c^+ \rightarrow J/\psi \tau^+ \nu $
 & 0.48\\
 $B_c^+ \rightarrow \psi^\prime e^+ \nu $
 & 0.132 \\
 $B_c^+ \rightarrow \psi^\prime \tau^+ \nu $
 & 0.011\\
 $B_c^+ \rightarrow  D^0 e^+ \nu $
 & 0.004 \\
 $B_c^+ \rightarrow  D^0 \tau^+ \nu $
 & 0.002 \\
 $B_c^+ \rightarrow  D^{*0} e^+ \nu  $
 & 0.018  \\
 $B_c^+ \rightarrow  D^{*0} \tau^+ \nu  $
 & 0.008 \\
 $B_c^+ \rightarrow  B^0_s e^+ \nu  $
 & 4.03  \\
 $B_c^+ \rightarrow B_s^{*0} e^+ \nu  $
 & 5.06 \\
  $B_c^+ \rightarrow B^0 e^+ \nu  $
 & 0.34\\
 $B_c^+ \rightarrow B^{*0} e^+ \nu  $
 & 0.58 \\
 $B_c^+ \rightarrow \eta_c \pi^+$
 & 0.20\\
 $B_c^+ \rightarrow \eta_c \rho^+$
 & 0.42\\
 $B_c^+ \rightarrow J/\psi \pi^+$
 & 0.13\\
 $B_c^+ \rightarrow J/\psi \rho^+$
 & 0.40\\
 $B_c^+ \rightarrow \eta_c K^+ $
 & 0.013\\
 $B_c^+ \rightarrow \eta_c K^{*+}$
 & 0.020\\
\hline
\end{tabular}
\begin{tabular}{|l|r|}
\hline
~~~~~Mode & BR, \%\\
\hline
 $B_c^+ \rightarrow J/\psi K^+$
 & 0.011\\
 $B_c \rightarrow J/\psi K^{*+}$
 & 0.022\\
 $B_c^+ \rightarrow D^+
\overline D^{\hspace{1pt}\raisebox{-1pt}{$\scriptscriptstyle 0$}}$
 & 0.0053\\
 $B_c^+ \rightarrow D^+
\overline D^{\hspace{1pt}\raisebox{-1pt}{$\scriptscriptstyle *0$}}$
 & 0.0075\\
 $B_c^+ \rightarrow  D^{\scriptscriptstyle *+}
\overline D^{\hspace{1pt}\raisebox{-1pt}{$\scriptscriptstyle 0$}}$
 & 0.0049\\
 $B_c^+ \rightarrow  D^{\scriptscriptstyle *+}
\overline D^{\hspace{1pt}\raisebox{-1pt}{$\scriptscriptstyle *0$}}$
 & 0.033\\
 $B_c^+ \rightarrow D_s^+ \overline
D^{\hspace{1pt}\raisebox{-1pt}{$\scriptscriptstyle 0$}}$
 & 0.00048\\
 $B_c^+ \rightarrow D_s^+
\overline D^{\hspace{1pt}\raisebox{-1pt}{$\scriptscriptstyle *0$}}$
 & 0.00071\\
 $B_c^+ \rightarrow  D_s^{\scriptscriptstyle *+} \overline
D^{\hspace{1pt}\raisebox{-1pt}{$\scriptscriptstyle 0$}}$
 & 0.00045\\
 $B_c^+ \rightarrow  D_s^{\scriptscriptstyle *+}
\overline D^{\hspace{1pt}\raisebox{-1pt}{$\scriptscriptstyle *0$}}$
 & 0.0026\\
 $B_c^+ \rightarrow \eta_c D_s^+$
 & 0.86\\
 $B_c^+ \rightarrow \eta_c D_s^{*+}$
 & 0.26\\
 $B_c^+ \rightarrow J/\psi D_s^+$
 & 0.17\\
 $B_c^+ \rightarrow J/\psi D_s^{*+}$
 & 1.97\\
 $B_c^+ \rightarrow \eta_c D^+$
 & 0.032\\
 $B_c^+ \rightarrow \eta_c D^{*+}$
 & 0.010\\
 $B_c^+ \rightarrow J/\psi D^+$
 & 0.009\\
 $B_c^+ \rightarrow J/\psi D^{*+}$
 & 0.074\\
 $B_c^+ \rightarrow B_s^0 \pi^+$
 & 16.4\\
 $B_c^+ \rightarrow B_s^0 \rho^+$
 & 7.2\\
 $B_c^+ \rightarrow B_s^{*0} \pi^+$
 & 6.5\\
 $B_c^+ \rightarrow B_s^{*0} \rho^+$
 & 20.2\\
\hline
\end{tabular}
\begin{tabular}{|l|r|}
\hline
~~~~~Mode & BR, \%\\
\hline
 $B_c^+ \rightarrow B_s^0 K^+$
 & 1.06\\
 $B_c^+ \rightarrow B_s^{*0} K^+$
 & 0.37\\
 $B_c^+ \rightarrow B_s^0 K^{*+}$
 & --\\
 $B_c^+ \rightarrow B_s^{*0} K^{*+}$
 & --\\
 $B_c^+ \rightarrow B^0 \pi^+$
 & 1.06\\
 $B_c^+ \rightarrow B^0 \rho^+$
 & 0.96\\
 $B_c^+ \rightarrow B^{*0} \pi^+$
 & 0.95\\
 $B_c^+ \rightarrow B^{*0} \rho^+$
 & 2.57\\
 $B_c^+ \rightarrow B^0 K^+$
 & 0.07\\
 $B_c^+ \rightarrow B^0 K^{*+}$
 & 0.015\\
 $B_c^+ \rightarrow B^{*0} K^+$
 & 0.055\\
 $B_c^+ \rightarrow B^{*0} K^{*+}$
 & 0.058\\
 $B_c^+ \rightarrow B^+ \overline{K^0}$
 & 1.98\\
 $B_c^+ \rightarrow B^+ \overline{K^{*0}}$
 & 0.43\\
 $B_c^+ \rightarrow B^{*+} \overline{K^0}$
 & 1.60\\
 $B_c^+ \rightarrow B^{*+} \overline{K^{*0}}$
 & 1.67\\
 $B_c^+ \rightarrow B^+ \pi^0$
 & 0.037\\
 $B_c^+ \rightarrow B^+ \rho^0$
 & 0.034\\
 $B_c^+ \rightarrow B^{*+} \pi^0$
 & 0.033\\
 $B_c^+ \rightarrow B^{*+} \rho^0$
 & 0.09\\
 $B_c^+ \rightarrow \tau^+ \nu_\tau$
 & 1.6\\
 $B_c^+ \rightarrow c \bar s$
 & 4.9\\
\hline
\end{tabular}
\end{center}
\caption{Branching fractions of exclusive $B_c$ decay modes \cite{R7}}\label{tab}
\end{table}

From table \ref{tab} it is clearly seen, that main modes of $B_c$-meson decay are connected with $B_c\to B_s$ transition. The branching of $B^+_c\to B^0_s\pi^+$ and $B_c\to B^0_s\rho$ decays are 16.4\% and  7.2\% respectively. The branching fractions of $B_c\to B_s^*\pi$ and $B_c\to B_s^* \rho$ are rather large also:
\begin{eqnarray*}
\mathrm{Br}(B_c\to B_s^*\pi) &=& 6.5\%,\quad \mathrm{Br}(B_c\to B_s^*\rho) = 20.2\%.
\end{eqnarray*}
The sum of these branching fractions gives $\sim 50\%$. Inspired by the recent progress in detection of $B_s$-mesons, used for measurement of $B_s$ oscillations, one could expect, that this will give new ways of $B_c$-meson detection through $B_c\to B_s$ decays \cite{R1,R2}, that are caused by $c$-quark decay.

\section{$B_c$ production}

Hadronic production of $B_c$-meson was considered in a number of theoretical works. These works can be divided into two groups:
\begin{enumerate}
\item articles, where all $O(\alpha_s^4)$ diagrams describing $B_c$ production were considered \cite{R4},
\item articles, where only diagrams with fragmentation of $b$-quark into $B$-meson were taken into account \cite{R10}.
\end{enumerate}
In \cite{R9} it was shown, that these approaches give the same results in the region of large transverse momentum $p_T(B_c)>p_T^0$. The value of the momentum $p_T^0$ depends on the quantum numbers of $B_c$-meson and varies from 30 to 40 GeV. For $p_T<p_T^0$ the fusion mechanism dominates. As a result the total contribution to $B_c$ production cross section (including the feeddown from the excited states) gives the value of order $10^{-3}$ of the cross section of $B$-meson production.

CDF and D0 \cite{R5,R6} collaborations give their results on $B_c$ production cross section ($\sigma(B_c)$) in the form of the ratio over the cross section of $B$-meson production ($\sigma(B)$):
\begin{eqnarray*}
R_e &=&\frac{\sigma_{B_c}\cdot B_r (B_c\to J/\psi e^+\nu)}{\sigma_B B_r (B\to J/\psi K^\pm)}
=0.282\pm 0.0038\pm 0.074
\end{eqnarray*}
in the kinematical region $p_T(B)> 4.0$ GeV and $|y(B)|< 1.0$. Similar result for $B_c\to J/\psi\mu^\pm \nu$ decay is
\begin{eqnarray*}
R_\mu =0.249\pm 0.045^{+0.107}_{-0.076}.
\end{eqnarray*}
We believe, that these results contradict theoretical estimates. Using known branching fractions $Br(B\to J/\psi K^{\pm})\simeq 1.10^{-3}$ and $Br(B_c\to J/\psi e^{\pm}\nu)=2\cdot 10^{-3}$, from table \ref{tab} one can see, that in this kinematical region the ratio
\begin{eqnarray*}
\frac{\sigma (B_c)}{\sigma (B)}
=
R_e
\frac
{Br(B\to J/\psi K^{\pm})Br B (b\to B^{\pm}}
{Br(B_c\to J/\psi e^{\pm}\nu)}
=\frac{0.282\cdot 10^{-3}\cdot 0.5}{2\cdot 10^{-2}}
=0.7\cdot 10^{-2},
\end{eqnarray*}
that is about an order of magnitude higher, then theoretical predictions.

Let us now return to detection of $B_c$ meson from the decay induced by the decay of $c$-quark (that is $B_c\to B_s$). If we rely on theoretical estimates of $B_c$-meson yield and experimental results of $B_s$ production cross section, that is suppressed by an order of magnitude in comparison with $B_{u,d}$, it is clear, that the ratio
\begin{eqnarray*}
\frac{\sigma_{B_c}}{\sigma_{B_s}}\sim 10^{-2}
\end{eqnarray*}
is an order of magnitude higher, than the same ratio for all $B$ mesons. If the branching fractions of the decay $B_c\to B_s+X$ is taken into account, from 5600 fully reconstructed $B_s$ events \cite{R1} we should expect 9 $B_c\to B_s\pi$ events and ~3 $B_c\to B_s\rho$ events. These numbers can be increased by an order of magnitude if partially reconstructed events with semileptonic $B_s$-decays are taken into account (CDF collaboration has detected 61 500 events of this type).

An important feature of cascade decays
\begin{eqnarray*}
  B_c &\to & B_s \to D_s
\end{eqnarray*}
is that in semileptonic decays leptons with equal charges are produced\footnote{
$B_s\leftrightarrow\bar B_s$ can, however, spoil this situation.
}. If hadronic decays $B_c\to B_s\pi$ or $B_c\to B_s\rho$ are detected, the charge of $\pi$ ($\rho$) meson coincides with the charge of the lepton produced in $B_s$ decay. Using the known values of semileptonic $B_s$ decay branching fractions and the ratio $10^{-2}$ of $B_c$ and $B_s$ production rates it is easy to estimate the number of $B_c$ mesons. For example, for $6\times 10^{-4}$ semileptonic $B_s$ decays we could expect $\sim 6\times 10^3$ initial $B_c$ mesons. Using the value $\mathrm{Br}(B_c\to B_s\pi)\approx 16\%$ we can expect $\sim 10^{3}$ decays $B_c\to B_s\pi$. Recalling that $B_s$ is observed in semileptonic decay, we obtain $\sim10^{2}$ events of this sort. The observation of these events will multiply the number of detected $B_c$ mesons many times. On the other hand, we will receive an additional opportunity to study the modes of $c$-quark decay inside $B_c$ meson.

Authors thank A.V. Luchinsky  and V.V. Kiselev for useful discussions. This work was partially supported by Russian Foundation for Basic Research under grant no.07-02-00417a.

\end{document}